\documentclass[%
 reprint,
 superscriptaddress,
 amsmath,amssymb,
 aps,
 pra,
 floatfix,
]{revtex4-2}
\usepackage{graphicx}
\usepackage{dcolumn}

\usepackage{bm}
\usepackage{amsmath,amssymb}
\usepackage[colorlinks=true,
            urlcolor=blue,linkcolor=blue,
            citecolor=blue,anchorcolor=blue,
            breaklinks=true]{hyperref}

\begin{document}

\preprint{APS/123-QED}

\title{Kinetics and Nucleation Dynamics in Ion-Seeded Atomic Clusters}

\author{M. G. Rozman}
 \affiliation{Department of Physics, University of Connecticut, Storrs, CT 06269, USA}
\author{M. Bredice}
 \email{mitchell.bredice@uconn.edu}
 \affiliation{Department of Physics, University of Connecticut, Storrs, CT 06269, USA}
\author{J. Smucker}
 \affiliation{Department of Physics, University of Connecticut, Storrs, CT 06269, USA}
\author{H. R. Sadeghpour}
\affiliation{%
 ITAMP, Center for Astrophysics $|$ Harvard \& Smithsonian, Cambridge, MA 02138, USA
}%
\author{D. Vrinceanu}
\affiliation{Department of Physics, Texas Southern University, Houston, TX 77004, USA}
\author{R. Cote}
   \affiliation{UMass-Boston, Department of Physics, Boston, MA 02125, USA}
\author{V. Kharchenko}
 \affiliation{Department of Physics, University of Connecticut, Storrs, CT 06269, USA}
 \affiliation{%
 ITAMP, Center for Astrophysics $|$ Harvard \& Smithsonian, Cambridge, MA 02138, USA
}%

\date{\today}

\begin{abstract}
The time-dependent kinetics of formation and evolution of nano-size atomic clusters is investigated and illustrated with the nucleation dynamics of ion-seed Ar$_n$H$^+$ particles. The rates of growth and degradation of Ar-atomic shells around the seed ion are inferred from Molecular Dynamics (MD) simulations. Simulations of cluster formation have been performed with accurate quantum-mechanical binary interaction potentials. Both the nonequilibrium and equilibrium growth of Ar$_n$H$^+$ are investigated at different temperature and densities of the atomic gas and seed ions. Formation of Ar$_{n\leq 40}$ shells is the main mechanism which regulates the kinetics of nano-cluster growth and the diffusive fluctuations of the cluster size distribution. The time-evolution of the cluster intrinsic energy and cluster size distributions are analyzed at the non-thermal, quasi-equilibrium, and thermal equilibrium stages of Ar$_n$H$^+$ formation. We've determined the self-consistent model parameters for the temporal fluctuations of the cluster size and found coefficients of the diffusive growth mechanism describing the equilibrium distribution of nano-clusters. Nucleation of haze and nano-dust particles in astrophysical and atmospheric ionized gases are discussed.
\end{abstract}
\maketitle
\section{Introduction}
The nature and characteristics of phase transitions in atomic and molecular systems strongly depend on the inter-particle interactions. These binary interactions determine how rapidly new phases form, and specifically regulate a nucleation of solid or liquid particles from gas and liquid phases. The Classical Nucleation Theory (CNT) \cite{LLkinetics, Zeldovich,ReviewCNT1,ReviewCNT2} was developed and successfully applied to the analysis of nucleation processes in macroscopic and submicron systems under conditions of thermal equilibrium. CNT predictions of the nucleation kinetics are essentially based on the Gibbs statistical rule and on stochastic (diffusive) formation of critical nucleation sizes\cite{LLkinetics, Zeldovich}. 

In contrast to CNT, theoretical modeling of nucleation and growth of nano-particles under local nonequilibrium and non-homogeneous conditions requires precise knowledge of inter-atomic forces and a detailed description of relaxation processes. We found that non-homogeneous nucleation of noble gas atoms, seeded by ions, is regulated via building up of ion-atom clusters from a sequence of stable atom shells. On the other hand, we observe large size fluctuations of nano-clusters under thermal equilibrium which resemble the diffusive growth of critical clusters in CNT models.
Nucleation of nano-size particles in ionized gases is an important process for a wide range of areas in physics, chemistry, astrophysics, and atmospheric sciences. Analysis of formation of nano-size haze and ice particles in the upper atmosphere of planets, satellites, exoplanets, and within debris disks is critically important for investigations of spectral properties of these atmospheres\cite{GRUNDY2018232,10.1093/mnras/stz2655,Gao2020,acp-18-4519-2018,Egorov2018,Protonated_water_exp} and circumstellar disks \cite{2000A&A...362.1127K,Dust_dist_Krivov,doi:10.1098/rsta.2016.0254}. Ions and charged nano-dust particles are major agents stimulating haze and ice cloud formations in the Earth upper atmosphere. The MD simulations, performed with accurate potentials of inter-particle interaction, can successfully describe both nonequilibrium and equilibrium processes of nucleation of nano-size clusters. Previous investigations were focused on the nucleation\cite{LJMD,ArMD}, structural properties\cite{PhysRevB.77.125434}, and phase transitions of pure argon clusters\cite{doi:10.1063/1.469470,REY1992273}. Meanwhile, there is a significant amount of nano-cluster research using supersonic beam experiments analyzing abundances of clusters with different numbers of atoms\cite{PhysRevA.98.022519,Schobel2011,Pure_He_Doped_Kr, Kuhn2016,PhysRevLett.51.1538,doi:10.1063/1.458275,PhysRevLett.53.2390,HARRIS1986316,MARK1987245,doi:10.1063/1.456898,doi:10.1063/1.457464}. The theoretical analyses of cluster structure, Ar$_n$H$^+$\cite{PhysRevA.98.022519, Ar_n_H+_QM,doi:10.1063/1.1485956,Ar2_3_structure} and small Lennard-Jones (LJ) clusters \cite{Vafayi2015}, have been concentrated on the explanation of different "magic numbers" obtained in different experiments. There is also significant work on small noble gas clusters focused on their fragmentation after ionization due to electron or proton impact \cite{C6CP07479K,PhysRevLett.99.083401,doi:10.1080/01442350701223045,doi:10.1063/1.2194552,Surdutovich2019}. Although nonequilibrium energy relaxation of excited clusters and following fragmentation involve the same binary potentials as cluster nucleation, the dynamics of these processes differ significantly. In our research, the formation and growth of nano-size Ar$_n$H$^+$ clusters is initiated by ionization of H atoms in the Ar and H gas mixture and is simulated with the use of the Large Atomic/Molecular Massively Parallel Simulator (LAMMPS) \cite{PLIMPTON19951}. The nucleation of Ar$_n$H$^+$ clusters have been investigated using the results of the MD simulations performed with quantum-mechanically calculated potentials and in the canonical (NVT) ensemble with Nos\'e-Hoover thermostat \cite{PLIMPTON19951}. The time-dependent kinetics of the nucleation and growth of critical Ar$_n$H$^+$ clusters and their approach to thermal equilibrium have been inferred from the results of our MD simulations. A role of ion seed particles in a phase transition has been analyzed in detail employing data of our simulations performed with different parameters of the Ar bath gas and densities of H$^+$ ions. 
 \section{Shell Structure, Energy, and Stability of Nano-Size Clusters }
 The accurate binary potentials of the Ar-Ar and Ar-H$^+$ interactions have already been used in our investigations of nucleation and growth of Ar$_n$H$^{+}_m$ solid or liquid particles; where large size clusters include many protons \cite{Oliver}. We briefly provide information on the major characteristics of these potentials. 
 \subsection{ Binary interaction potentials}
 
\begin{figure}[ht]
 \includegraphics[scale=0.28125]{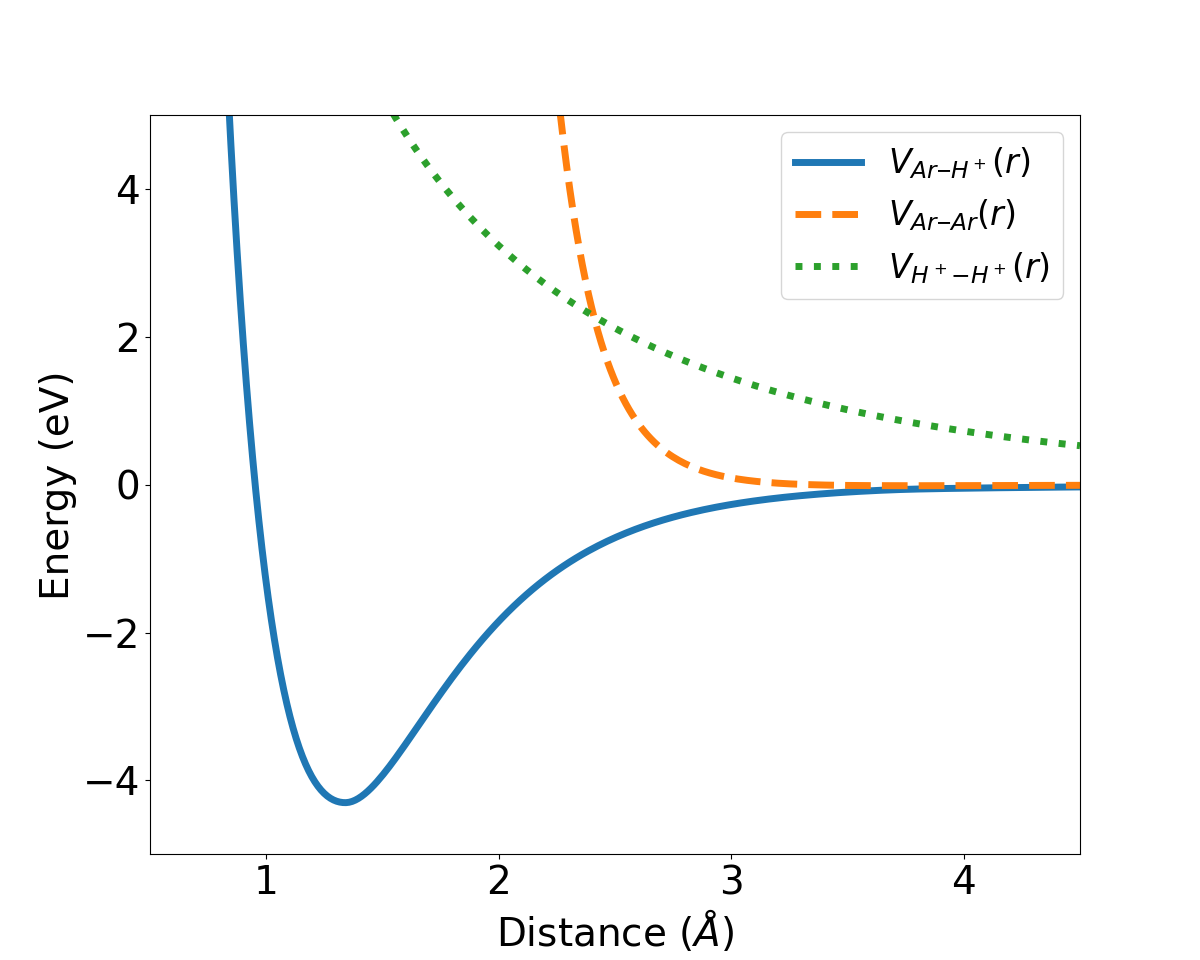}
 \caption{\label{fig:Pot}Potential energy curves for the MD simulations. The screened H$^+$-H$^+$ interaction is Debye shielded and modeled as a Yukawa-type potential, due to the electron and proton fields. The Ar-Ar interaction is modeled by a LJ potential and describes the short-ranged vdW interaction between Ar atoms. The Ar-H$^+$ interaction asymptotically leads to a r$^{-4}$ polarization potential and is the longest range attraction in our simulation. The repulsion and attraction energies of the Ar-Ar and Ar-H$^+$ interactions are approximately equalized at $\sim$ 2.5\AA.}
 \end{figure}

 The comprehensive description of the binary interaction potentials used in our simulations is given in \cite{Oliver}. To summarize the details, the Ar-Ar binary interaction is a LJ 6-12 short-range van der Waals potential with a well depth of 0.012 eV and an equilibrium atomic separation of 3.75 \AA. The Ar-H$^+$ interaction at its deepest is more than 4 eV and it asymptotically leads to a r$^{-4}$ polarization potential. This interaction is much stronger than the Ar-Ar interaction due to the ion Coulomb field polarizing the neutral atom. Thus, Ar-H$^+$ potential creates the bulk of the inter-particle attraction in our simulations, while the Ar-Ar interaction does not substantially contribute to any attraction forces in our simulations of Ar$_n$H$^+$ cluster growth with n $\leq$ 40. On the other hand, the short range Ar-Ar repulsion together with Ar-H$^+$ potential controls parameters of the Ar-shells. The ion-ion interaction is Debye shielded and is modeled as a Yukawa-type interaction in LAMMPS. A sketch of the potentials is given in Fig.\ref{fig:Pot}. All potentials used in our modeling do not include explicit three-body or higher many-body contributions.

\subsection{Geometrical structures and cluster energies: emergence of primitive symmetry structures.}
 
  Nucleation of nano-size clusters has been studied in detail for neutral noble gases \cite{Berry}. Neutral atoms at low temperatures form cluster shells minimizing the cluster energy. Larger cluster shells with increasing numbers of atoms can be transformed to regular crystal structures, if the reservoir temperature is below the melting point. Stability of nano-cluster shells and their geometrical parameters are sensitive to characteristics of inter-atomic potentials \cite{Berry}. Nucleation of atoms seeded by ions or charged nano-particles differs significantly from cluster formation in homogeneous neutral gases. The strong attraction force between a seed ion and Ar atoms dominates the dynamics during the initial stages of Ar$_n$H$^+$ formation. This ionic force creates tighter confinement of the Ar$_n$H$^+$ cluster with larger binding energies for Ar atoms in close proximity to the H$^+$. The distance between Ar atoms and ion, i.e. the shell radius $\overline{r}$, is the same for all atoms belonging to the same shell, and the shell radius $\overline{r}$ practically does not depend on the number of atoms in the shell. The number of Ar atoms $n_s$ in the specific shell is restricted due to conditions minimizing a cluster potential energy. The maximal value $n_s$ occurs for the closed shells with $n_s$=4,6, .... The specific geometrical configurations, tetrahedral and octahedral, are the primitive symmetry configurations around the central seed ion that minimize the potential energy $U(n)$ of clusters with closed shells. The primitive symmetry of the first Ar atomic shell, tetrahedral, disagrees with the structure found in the literature \cite{PhysRevA.98.022519, Ar_n_H+_QM,doi:10.1063/1.1485956,Ar2_3_structure}, which has the innermost layer of the cluster being a linear Ar-H$^+$-Ar cluster. Nevertheless, we expect that simulations of the cluster growth and cluster size distributions cannot be affected strongly by the semiclassical approximation used in our MD model. This is because in the both semi-classical and quantum-mechanical models, the cluster growth of Ar$_{n-1}$H$^+$ + Ar $\rightarrow$ Ar$_n$H$^+$ occurs via captures of Ar atoms into highly excited configuration states with a following relaxation of the cluster intrinsic energy to the ground states. The energies, $U(n)$, which correspond to the minimal total energies at zero temperature, are given in Table 1 together with an indication of the symmetry of the cluster configurations (shells), Ar-H$^+$ inter-atomic distances $\overline{r}$, and shortest Ar-Ar distances within each shell that provide these minima. 
  \begin{table*}[htb]
  \begin{ruledtabular}
  \label{tbl:clusters}
  \begin{tabular}{rlllr}
  \textbf{Cluster}
  &
   \textbf{Symbol}   
  &
   \textbf{$\bar{r}$ (\AA)}   
  &
  \textbf{r$_{Ar-Ar}$ (\AA)}
  &
   \textbf{$U(n)$ (eV)} \\
   \colrule
  Ar$_4$H$^+$  & 1T$^4$       & 1.69    & 2.76  & -10.2333 
  \\ 
  Ar$_8$H$^+$  & 1T$^4$2T$^4$    & 1.69, 3.57 & 2.76, 5.83    & -10.6406
  \\ 
  Ar$_{14}$H$^+$ & 1T$^4$2T$^4$3O$^6$ & 1.69, 3.58, 4.29 & 2.76, 5.84, 6.07      & -11.1789
  \\ 
 \end{tabular}
 \end{ruledtabular}
 \caption{The minimal potential energy $U(n)$, average radii $\overline{r}$, and shortest Ar-Ar (within each shell) are shown for the cluster structure with closed cluster shells 1T$^4$ with the number of Ar atoms n$_S$=4, 1T$^4$2T$^4$ with n$_S$=8, and 1T$^4$2T$^4$3O$^6$ with n$_S$=14. The shell symmetries are contained in the notation, T = Tetrahedral, O = Octahedral. The shell radius $\overline{r}$ practically does not depend on the number of atoms in the specific shell and has same values for closed and open shells. Details of geometrical structure of the small nano-size clusters and their minimal energies can be found in \url{https://www.phys.uconn.edu/\~kharchenko/clusters/}}.
\end{table*}
  
\begin{figure}
\includegraphics[scale=0.26367]{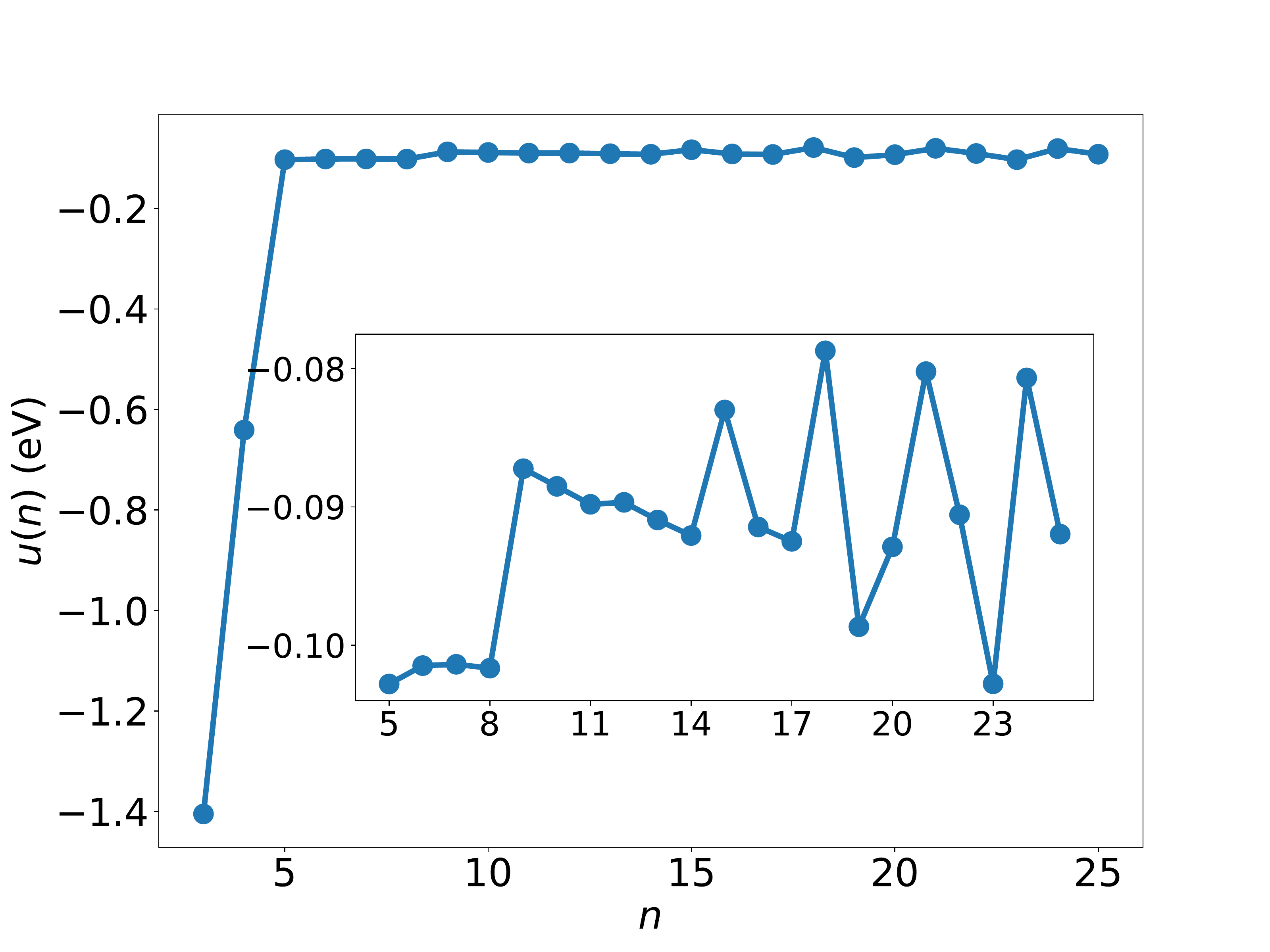}
\caption{\label{fig:ChemPotentials_Misha}The energies $u(n)$ of Ar-particle in the cluster shells, $u(n) = U(n)-U(n-1) =dU/dn$, as a function of the number of Ar atoms $n$ in the cluster. The minimal energy required at T=0 to remove an Ar atom from the Ar$_{n}$H$^+$ cluster is $|u(n)|$. The sharp increase of $u(n)$ function at $n \leq 4 $ corresponds to construction of the deepest 1T Ar shell. The plateau at n $\geq$ 5 shows that shell energies $u(n)$ of larger clusters are weakly sensitive to the value of cluster size n. The energy $|u(n)|$ is the maximal energy which can be transferred to the thermostat in Ar$_{n-1}$H$^+$ +Ar $\rightarrow$ Ar$_n$H$^+$ transitions. The irregular oscillations of $u(n)$, shown in the insert, reflect energy $n$-alternation of Ar atoms in the open shells. The value of these oscillations is reduced at  $n \gg 1$. }
\end{figure}

  The Ar atoms from the same shell have identical parameters, such as binding energy or an averaged distance from the cluster ion $\overline{r}$ . The minimal energies $u(n)$ of Ar atoms in the cluster shell, $u(n) = U(n) - U(n-1)=dU/dn$, were calculated for small Ar$_{n < 25}$H$^+$ clusters and results are shown in Fig.\ref{fig:ChemPotentials_Misha} as a function on $n$. The detachment energy required to remove an Ar atom from a Ar$_{n}$H$^+$ cluster is $|u(n)|$. The values of $u(n)$ take into account the mean potential field created by all cluster Ar atoms and the H$^+$. The single particle energy $u(n)$ plays the role of the chemical potential at $T=0$. The value of $u(n)$ at n$\leq$4 rises sharply with $n$ and then plateaus at $n \geq$4. The four Ar atoms form the stable tetrahedral shell of atoms closest to the seed ion. The energy required to remove any of the Ar atoms from the Ar$_4$H$^+$ shell at T=0 is $u(n=4)\simeq $ 0.64 eV that corresponds to the temperature $\sim$ 7300 K. This energy is significantly higher than thermal energies considered here, and the Ar$_{4}$H$^+$ clusters are therefore stable in our simulations. A detachment of an Ar atom from the next tetrahedral shell (2T$^4$ in Table.1) requires 0.1 eV$\sim 800$ K. This shell should be stable at low temperatures (T$\sim$90 K) but can be depopulated under thermal equilibrium at T$\sim$200 K at sufficiently dilute Ar densities \cite{LLStatPhys}.
  The ion field removes the diffusion barrier \cite{Zeldovich,ReviewCNT1} of nucleation for small clusters and accelerates their nucleation. The charge-induced growth of small clusters occurs via the capture of free Ar atoms into cluster shells. The growth of Ar$_{n}$H$^+$ particles is restricted by thermal detachment of Ar atoms from cluster shells or by formation of new phases; for example, the stable Ar$_{n}$H$_m^+$ large clusters or crystals at low temperatures \cite{Oliver}. In our current simulations, the stable cluster shells with large number of Ar atoms were not observed at low densities or high temperatures. This is because the rate of evaporation of cluster-bound Ar atoms increases under these conditions significantly with respect to the shell growth rate causing depopulation of outer shells.\\

\section {\texorpdfstring{Methods and Results of MD Simulations of $Ar_nH^+$ Cluster Formation}{Section: Methods and Results of MD Simulation}}
Each MD simulation begins by randomly generating coordinates for the atoms and ions; the atoms and ions are separated by a minimum of 3 \AA. The simulations are run at temperatures, $90 < T < 600$K, with a variable number of ions from 0 to 200, and contain a fixed number of Ar atoms (1000) in the simulation box. Ar atom velocities are initialized with the Maxwell-Boltzmann distribution appropriate to the selected temperature, using LAMMPS built in "create velocity" function, but the initial velocity of the H$^+$ ions is assigned to be 0. To achieve different densities of Ar atoms and H$^+$ ions the size of the simulation box is adjusted, with the majority of the simulations performed at the Ar atom density of $10^{20}$ cm$^{-3}$. This density is chosen to represent a dense gas and allows for a faster convergence of the cluster growth. All simulations are run using the canonical ensembles (NVT) with the Nose-Hoover thermostat function contained in LAMMPS. The time step for our simulations was 1 fs and the thermostat temperature damping timescale was 100 fs to avoid sharp changes of the kinetic energies of atomic particles.
\subsection{Cluster Definition} 
The method to extract the clusters from the results of simulations is done in two stages, by first a geometrical selection of groups of close atoms, clusters, using the DBscan algorithm\cite{dbscan96}, with its implementation in the Julia programming language through the {\it Clustering.jl} package, and next computation of the cluster intrinsic energies to verify that all atoms in these clusters are bound. 

{\it Stage 1:} The DBscan algorithm searches for neighbors around every atom in the simulation, and if another atom is within a defined cutoff distance $d$ then it is selected to be a part of a possible geometric Ar$_n$H$^+$ cluster. In this first step we only search for clusters that include a single ion and use a cutoff distance of $d$=4 \AA. This cutoff was chosen from the energy minimization calculations, to ensure that the outer Ar shells with $n<40$ can be detected. The geometric cutoff selection could include "false" selections, e.g. a group of atoms and ions which are accidentally close to each other in one specific snapshot of the simulation, but not the next. Such groups of unbound particles are dissolved in a short time. These candidates for the Ar$_n$H$^+$ clusters can be analyzed and rejected on the second stage of the cluster-verification process.

{\it Stage 2:} We implement another step to our definition, that the cluster total energy in the Center of Mass (CM) frame must be negative, since this would indicate a truly bound system. The Stage 2 verification yields values of atomic binding energies; this allows the comparison to the minimal cluster energies obtained in minimization calculations for each cluster size (Fig.\ref{fig:ChemPotentials_Misha}). The irregular n-variations of $u(n)$ reflect the reconstruction of cluster shells with the addition of new atoms. The n-variations of the optimal configuration of cluster shells have been established in classical and quantum calculations of atomic binding energies \cite{PhysRevA.98.022519, Ar_n_H+_QM, doi:10.1063/1.1485956}. These variations should vanish at “n $\rightarrow$ $\infty$” when $u(n)$ provides the minimal potential energy of atomic particle on the surface of a macroscopic Ar crystal at T=0. The real total potential energies for clusters obtained in our simulations are slightly larger than the minimum  potential energy. The intrinsic cluster energy obtained in simulations is typically a few percent higher than the minimal potential energy. This excess of the intrinsic energy arises from the vibration of Ar atoms in the cluster (the kinetic energy of thermal motion), since our minimal potential energy is calculated at T=0. 
 \subsection{Simulations of the Nucleation Kinetics}
 We have performed a set of MD simulations of the Ar$_n$H$^+$ cluster growth at different temperatures and concentrations of Ar atoms and charged particles H$^+$. Nascent H$^+$ ions produced at t=0 in the thermal Ar gas quickly become centers of cluster nucleation. The formation of first atomic shells around charge centers is accompanied by a local release of significant energy. This initial stage of the nucleation process can be described as a non-thermal phase of the cluster formation. The large fraction of released energy is transferred to light particles, the protons, and later distributed between atomic particles and absorbed by the LAMMPS thermostat. Growth of cluster shells leads to diminishing binding energies of Ar atoms in Ar$_n$H$^+$ clusters. Thus, the energy release is reduced with cluster growth and the entire system (Ar gas and Ar$_n$H$^+$ clusters) approaches thermal equilibrium. 
 
 The MD simulations provide near complete information about spatial and velocity distributions of all atomic particles in the free Ar gas and in Ar$_n$H$^+$ clusters. Every time interval $\delta t=$ 20 ps, data is dumped for the entire simulation.This value of $\delta t$ allows to observe different phase space configurations at considered densities and temperatures of atomic particles. Analysis and calculations of physical and geometrical parameters are performed using these 20 ps-snapshots. 
\subsection{\texorpdfstring{Cluster formation and velocity distributions of Ar and H$^+$ particles}{Subsection: Cluster Formation and Velocity Distributions}}
 The initial stages of cluster growth occur in a nonequilibrium regime, lasting for the first $\sim$ 1-5 ns. The duration of this nonequilibrium stage depends on the initial conditions, the Ar gas parameters, and density of H$^+$ ions. This is apparent from the analysis of Ar and H$^+$ velocity distributions. We see that after the first few steps (each step is 20 ps long), the Ar and H$^+$ velocity distributions are non-Maxwellian. In each case the majority of the atoms/ions have small velocities, but with long high-velocity tails. The energy source for the hot particles in high-velocity tails is the formation of small clusters: capture of free Ar atoms into an open Ar$_n$H$^+$ cluster shells with $n \leq 8 $ is an exothermic process with a large energy release, as shown in Fig.\ref{fig:ChemPotentials_Misha}. Energy release decreases with growth of outer cluster shells and becomes comparable with values of typical thermal energy $kT$. 

\begin{figure}[ht!]
\includegraphics[scale=0.263]{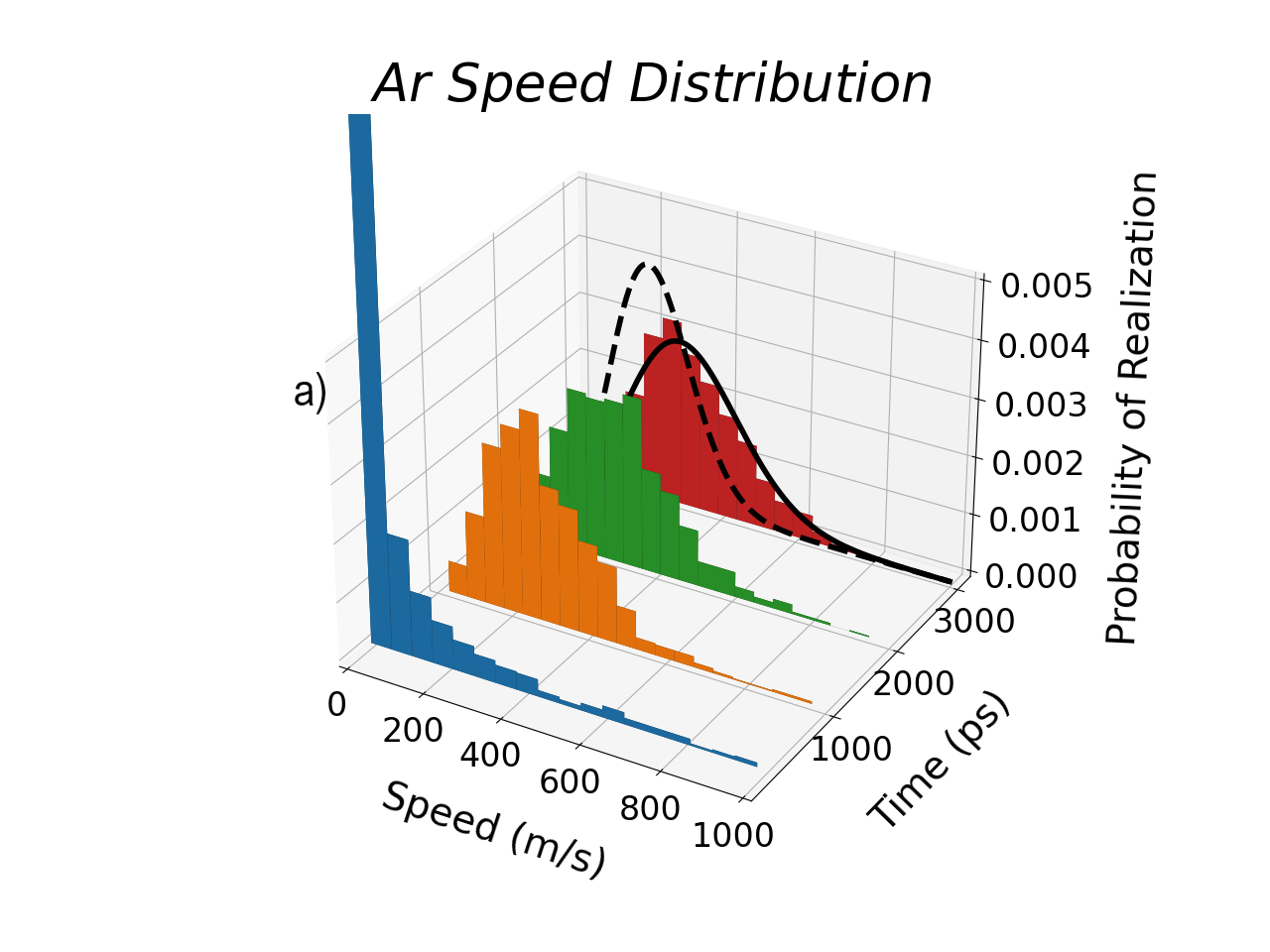}
\includegraphics[scale=0.263]{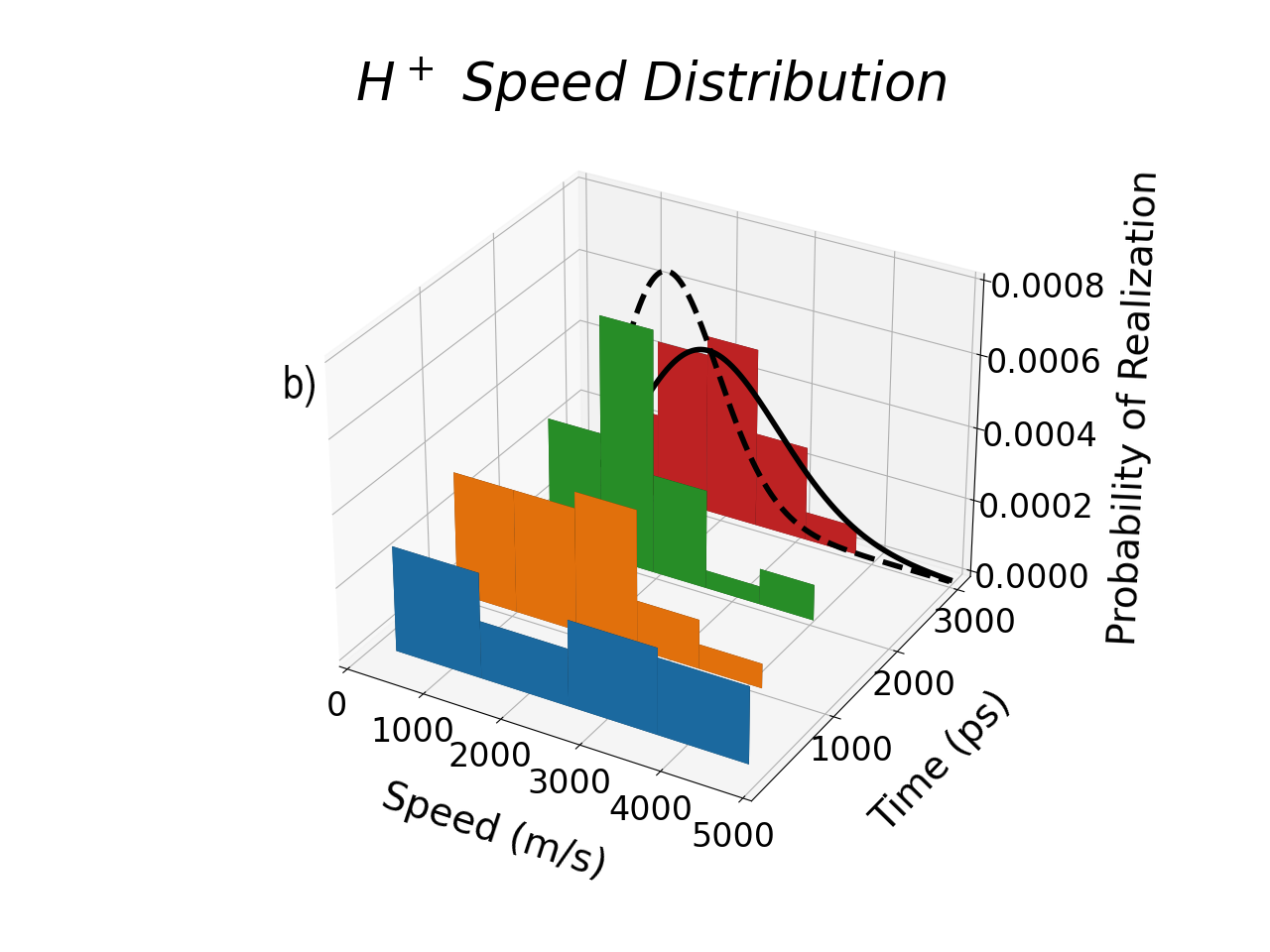}
\caption{\label{fig:Speed_Dist}The time-evolution of Ar gas velocity distributions, a), and H$^+$ ions, b), during growth of Ar$_n$H$^+$ clusters at 200K. The Ar atom density is $10^{20}$cm$^{-3}$ with 1000 Ar atoms and 60 H$^+$ ions in the simulation box. The small fraction of Ar atoms/H$^+$ ions in the high velocity tail are omitted. The distributions are normalized to the unity. The theoretical Maxwell-Boltzmann velocity distributions are shown for 100K (dashed line), and 200K (solid line) together with results of simulations at 3 ns. These figures illustrate the time scale required for the relaxation of velocity distribution functions.}
\end{figure}

As the simulation progresses, the peak of the velocity distribution shifts until the distribution becomes nearly Maxwellian. Although an insignificant tail of high velocities persists in the $v$-distributions for the entire duration of the simulations, the bulk of the atoms/ions remain in the thermalized Maxwellian-type distribution. The time-evolution of Ar and H$^+$ velocity distribution functions are show in Fig.\ref{fig:Speed_Dist}a and Fig.\ref{fig:Speed_Dist}b. The argon atoms appear cold at 200 ps due to the large energy losses: Ar atoms transfer their kinetic energy to the initially "frozen" light particles (the ions) at the beginning of the simulation. This specific aspect of the velocity relaxation does not influence formation of the first Ar$_n$H$^+$ shell, because atomic binding energies in the cluster shell are roughly two orders of magnitude larger than thermal energies.

\subsection{Nonequilibrium and Equilibrium Stages of Cluster Nucleation and Growth }
\subsubsection{\texorpdfstring{Stages of Ar$_n$H$^+$ cluster formation}{Subsection:Stages of Cluster Formation}}
Analysis of the results of MD simulations shows at least three different phases in a formation of nano-size Ar$_n$H$^+$ clusters. In our cluster growth simulations, all small clusters with $ n \leq 40 $ have distinct shell structures minimising their potential energies.

\begin{figure}
\includegraphics[scale=0.26367]{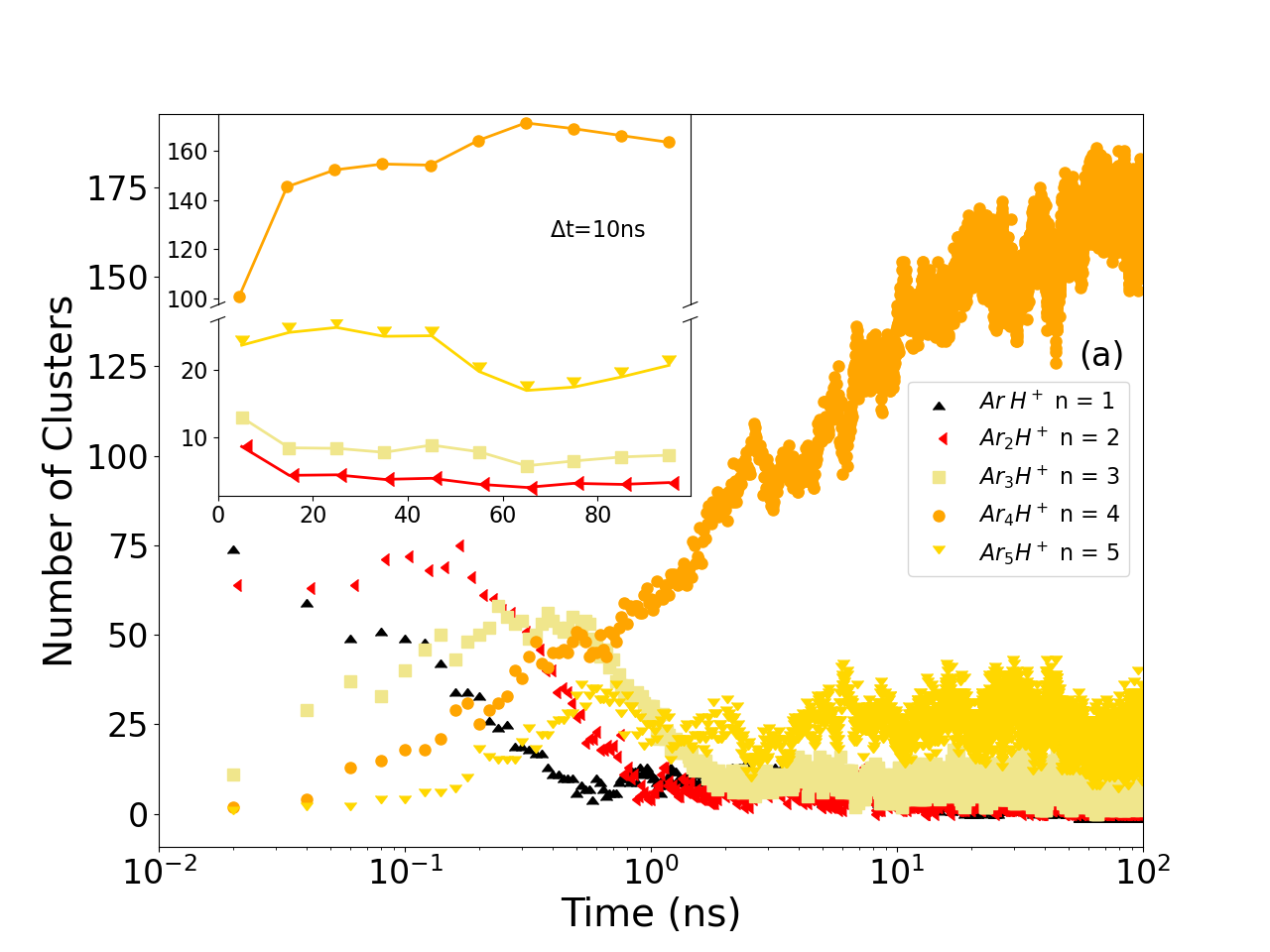}
\includegraphics[scale=0.26367]{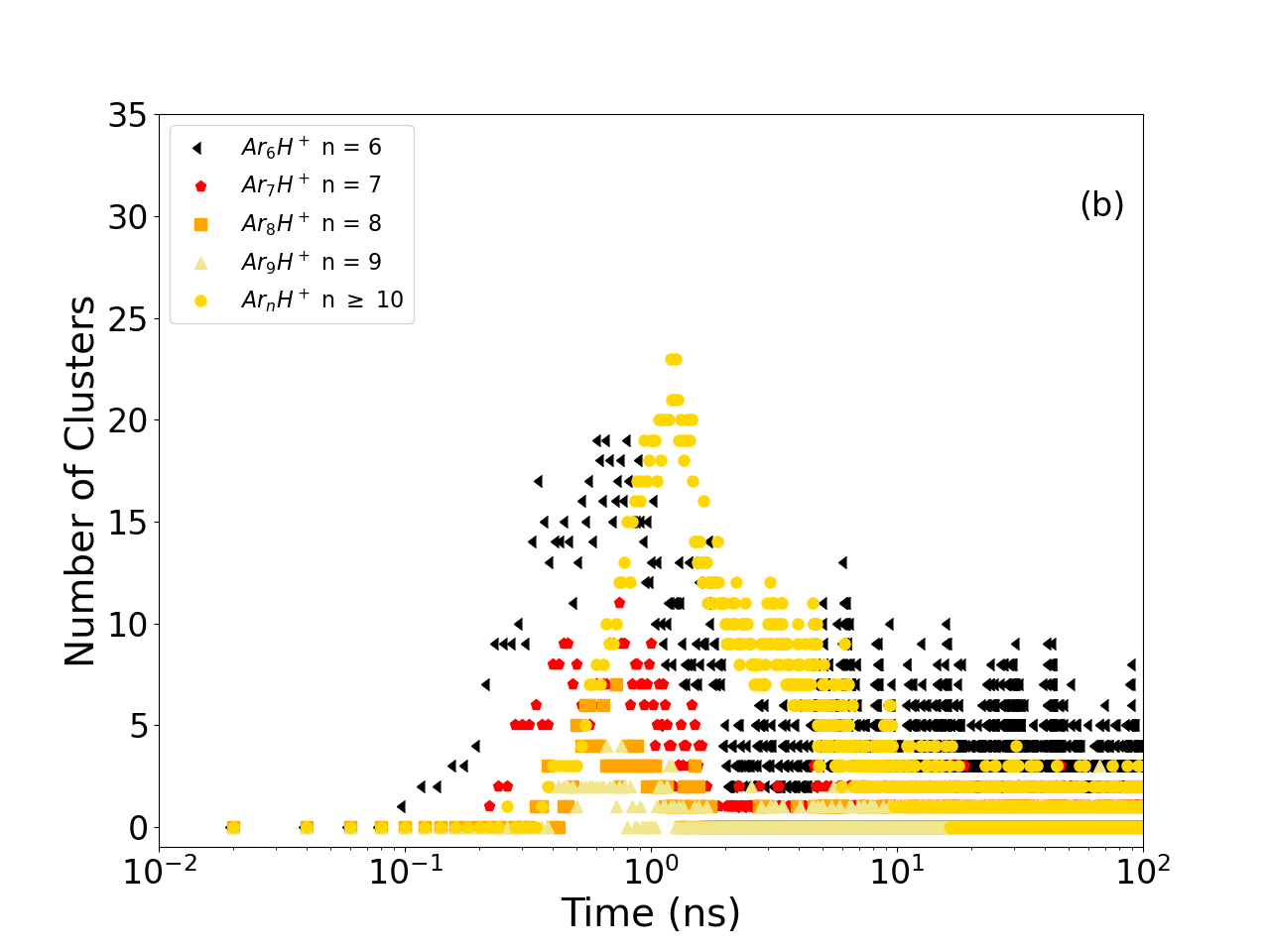}
\caption{\label{fig:VasFig1}The abundances $N_c(n,t)$ of Ar$_n$H$^{+}$ clusters with different number of Ar atoms n ($n \leq 5$ in Fig.4a, $ 6\leq n \leq 9$ and the cumulative abundance of n $\geq$ 10 in Fig.4b) as functions of time $t$. Clusters are formed by an ensemble of 200 H$^+$ ions embedded into the equilibrium Ar gas at the temperature T= 200 K and the atom density of 10$^{20}$ cm$^{-3}$. H atoms have been ionised at the time $t$=0.The inset axes in Fig.4a show the abundances $N_c(n,t)$ of Ar$_n$H$^{+}$ ($2 \leq n \leq 5$) averaged over the interval $\Delta t=10ns$.}
\end{figure}

 The growth of the cluster size $n$ corresponds to a consequent filling of unoccupied Ar$_n$H$^+$ shells described in Section 2B. In Figs.\ref{fig:VasFig1}a and \ref{fig:VasFig1}b, we show an example of the Ar$_n$H$^{+}$ growth and degradation for the time $t\leq$ 100 ns after 200 H atoms have been ionized and mixed with the Ar gas.
 The numbers of clusters $N_c(n,t)$ with the specified number of Ar atoms $n$ are shown as functions of the time {\it t} with the regular snapshot interval of 20 ps. Simulations have been carried at the Ar atom density of 10$^{20}$ cm$^{-3}$ and the temperature T= 200K. In Figs.\ref{fig:VasFig1}a and \ref{fig:VasFig1}b; we can identify three distinct phases of nano-particle nucleation and growth. The first phase of nucleation is:
{\it (a) Nonequilibrium nucleation ($t \lesssim $ 2 ns )}. 

During this stage, Ar atoms are "captured" into the deepest shells. Protons create a strong potential field, and the process of capturing Ar atoms into closest cluster shells releases energies comparable with $u(n)$. Thermal energy fluctuations cannot detach Ar atoms from the first deeply bound shell (Fig.\ref{fig:ChemPotentials_Misha}). The stage (a) is an irreversible nonequilibrium process of the formation of the inner atomic shells of Ar$_n$H$^+$. The reduction of small cluster population $N_c(n,t)$ for $n=1-3$ shown in Fig.\ref{fig:VasFig1}a is explained by a capture of free Ar atoms into the closed cluster shells 1T$^4$. Outer cluster shells are formed at around $t\sim$0.1 - 1 ns. The cluster nucleation dynamics can be illustrated by the time-dependent abundance of Ar$_{n}$H$^+$ ($1 \leq n \leq 10$) clusters shown in Figs.\ref{fig:VasFig1}a and \ref{fig:VasFig1}b. 
At large $t$, non-significant populations of larger clusters with $n\geq 5$ reflect an efficient thermal evaporation of Ar atoms from outer cluster shells at T=200 K. Ar$_{5}$H$^+$ clusters are more abundant than other states of 2T$^n$ shells due to the $n-$diffusive behaviour of the relatively stable Ar$_{4}$H$^+$ clusters. Details of the diffusive regime will be described in the Section IV. 

{\it (b) Quasi-equilibrium growth ( 2 ns $\lesssim t \lesssim$ 60 ns ).} 

 The second stage  in the interval between $2$ ns $<t< 60$ ns, is characterized by distinct fluctuations of the cluster size shown in Fig.\ref{fig:VasFig1}a. During this quasi-equilibrium stage of the nucleation kinetics, the tetrahedral Ar$_4$H$^{+}$ clusters became the most abundant particles. The fluctuations of the cluster abundances relate to captures and losses of Ar atoms into specific cluster shells. Averaging of these fluctuations over statistically significant time-intervals with values between 10 ns and 20 ns, yields the smooth functions describing the average number of clusters $\overline{N}_c(n,t)$ with $n$ Ar atoms. The number of clusters depicted in Fig.\ref{fig:VasFig1}a have been averaged over statistically significant interval $\Delta$t=10ns and obtained results for $\overline{N}_{c}(n,t)$ are shown in Fig.\ref{fig:VasFig1}a's inset axes. The steady growth of $\overline{N}_c(n=4,t)$ of tetrahedral clusters during the stage (b) can be considered as a quasi-equilibrium nucleation and growth of a new phase.
 
{\it(c) Thermal equilibrium growth and size evolution ($ t \gtrsim $ 60 ns).}

The growth of $\overline{N}_c(n=4,t)$ is stopped in the stage (c) at $t\geq$ 60 ns, when the system $\{${\it free Ar atoms + Ar$_n$H$^+$ clusters} $\}$ has reached the steady state regime in cluster nucleation and evaporation, i.e. the system has relaxed to the state of full thermal equilibrium between different phases (Fig.\ref{fig:VasFig1}a). The averaged number of Ar$_n$H$^+$ remain constant for the entire duration of stage (c) 60 ns $< t <$ 100 ns, and typical fluctuations of the $n$-number of Ar atoms bound by H$^+$ ions are described by thermal fluctuations.
The time boundaries for these stages, in Figs.\ref{fig:VasFig1}a and \ref{fig:VasFig1}b, are given as approximates; since the actual time-boundaries between stages depend on Ar and H$^+$ densities and temperatures used in each specific simulation.
\subsubsection{\texorpdfstring{Dynamics of the independent growth of Ar$_n$H$^{+}$ clusters}{Subsection: Dynamics of Independent Growth}}
\begin{figure}
\includegraphics[scale=0.325]{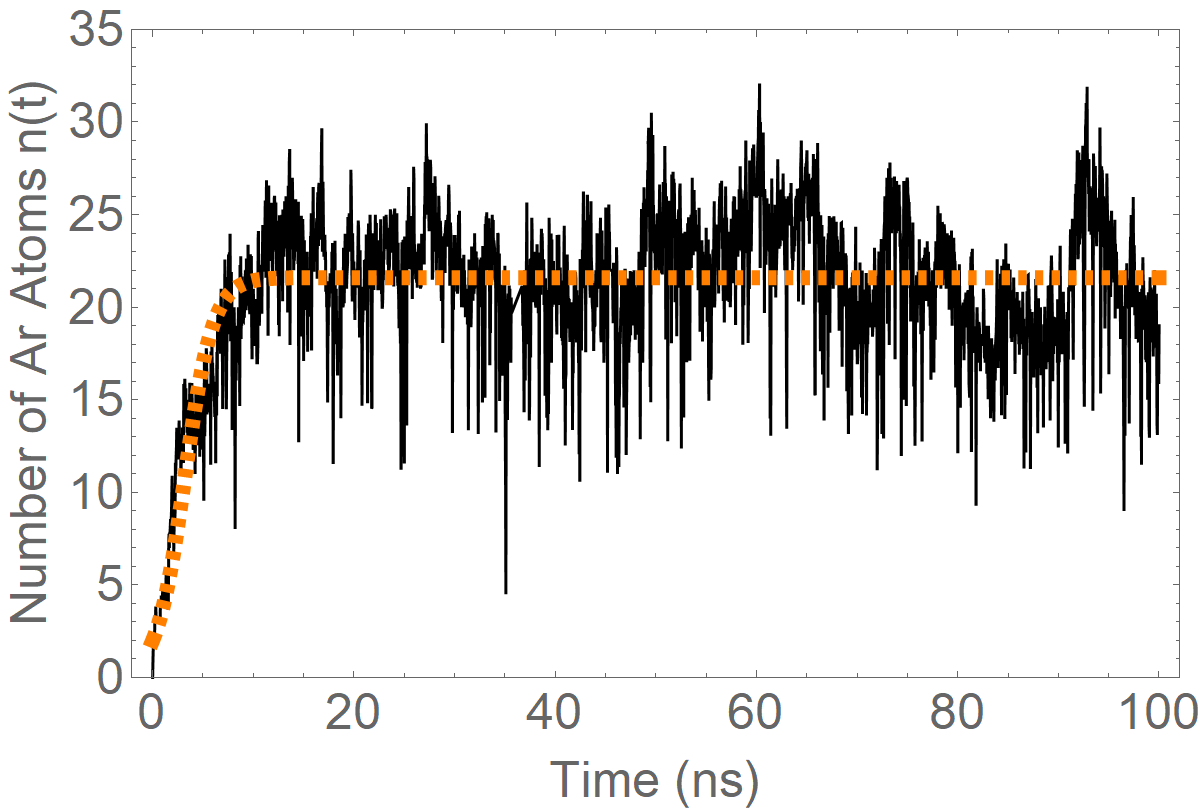}
\caption{\label{fig:single100ns90k}MD simulations of number of Ar atoms $n(t)$ in a single Ar$_{n(t)}$H$^+$ cluster. The simulation time is t= 100 ns and the Ar gas temperature is T= 90 K and the number of Ar atoms in the simulation box N=10$^3$. The dashed line is the theoretical curve for an average cluster size $\overline{n}(t)$ as a function of time. The time delay parameter $t_d= 3.2$ ns and the scaling relaxation time is $t_T= 1.4$ ns.}
\end{figure}

 Under conditions of the independent growth of Ar$_n$H$^{+}$ clusters, the averaged time-dependent characteristics of a single cluster describe parameters of the ensemble of independent protons. This ergodic statement is not valid for interacting clusters, when consolidation of Ar$_n$H$^+$ clusters leads to formation of a new phase like large Ar$_n$H$_m^+$ nano-crystals or liquid droplets \cite{Oliver}.
The nucleation and growth of independent clusters have been simulated for a single H$^+$ ion at different temperatures and at the constant number of Ar atoms $N=$10$^3$ in the simulation box. Nucleation of the smallest Ar$_n$H$^+$ clusters and consequent growth of Ar shells have been studied up to $t=100$ ns. In Fig.\ref{fig:single100ns90k}, the actual time-evolution of the cluster size $n(t)$ is shown for a single Ar$_{n(t)}$H$^+$ cluster at T= 90 K. 
The dashed red curve shows the theoretical data on the cluster growth based on a simplified two state approximation \cite{TwoStates}. In this model, the shell growth and cluster size oscillations are described by transitions between two states: the state of free Ar atoms in the simulation box of volume $V$ and the bound state of Ar atoms in the outer shell of the cluster of size $\overline{n}(t)$. We only consider the binding energies $\epsilon(n)$ of Ar atoms in the outer cluster shells and substitute instead $n(t)$ its values averaged over fluctuations: $n \rightarrow \overline{n}(t)$. The $\epsilon(n)$ energy in a specific shell depends on a mean-field created by H$^+$ and all Ar atoms attached to the ion. 
The self-consistent energy $u(n)$ is an essential part of the binding energy $\epsilon(n)$, though $\epsilon(n)$ may include contributions of different shell configurations mixed by thermal fluctuations. We assume, for simplicity, that cluster growth occurs in subsequent captures/detachments of a single Ar atom $ \overline{n}(t) \rightarrow \overline{n}(t) \pm 1$ and the number of Ar atoms in the gas is large $N \gg\overline{n}(t)$. For the two states model, each Ar atom can occupy the outer shell of the Ar$_{\overline{n}(t)}$H$^+$ cluster or stays in the free gas state. The partition function $Z$ of an ensemble of Ar atoms can be written as:
 \begin{equation}
   Z= \frac{z_{Ar}^N}{N!} ,~~~~ z_{Ar}= V/\lambda_T^3 + g(\overline{n}(t)) \exp\left[-\frac{\epsilon(\overline{n}(t))}{kT}\right],
 \end{equation}
 where $z_{Ar}$ is the partition function for a Ar atom under condition of the Maxwell-Boltzamnn statistics, and $\lambda_T= \sqrt{2 \pi \hbar^2/m kT } $ is the thermal de-Broglie wave of Ar atoms. The first term in the expression for $z_{Ar}$ corresponds to the occupied number of states in the classical gas with the fixed temperature and  particles density ($N \gg\overline{n}(t)$), and the second one corresponds to the contribution of the outer cluster shell with  the energy $\epsilon(\overline{n}(t))$  and statistical weight $g(\overline{n}(t))$. The values of $g(\overline{n}(t))$ are approximately proportional to the product of geometrical cluster volume $ V_{c}(\overline{n}(t))$ and the thermal momentum space volume: $g(\overline{n}(t)) \sim V_{c}(\overline{n}(t))/ \lambda_T^3 $. The presence of $N_{H^+}$ independent H$^+$ ions in the simulation box could increase the cluster statistical weight by $N_{H^+}$ times.
 
 The two-state approximation describes the populations of these states via the effective chemical potential $\mu_{eff}$, defined  for Ar atoms  in the outer cluster shell. The  $\mu_{eff}$-value takes into account different statistical weights of the gas and cluster states \cite{TwoStates}. Our simulations include the strong nonequilibrium initial condition: all deep cluster shells are empty at $t=0$. Under this condition, the value of $\mu_{eff}$, regulating population of the cluster shells should depend on time: $\mu_{eff}(t)= \mu_{eff}(N,T,t)$. This effective chemical potential will asymptotically approach the equilibrium value $\mu(N,T)$ when the relaxation processes are accomplished. The population of cluster shells $\overline{n}(t)$ can be expressed as:
 \begin{equation}
   \overline{n}(t) = N \frac{\exp\left[-\Delta(t,T)\right]}{1+ \exp\left[-\Delta(t,T)\right]} \simeq n_c ~ \frac{\exp\left[\frac{(t- t_d)}{t_T}\right]}{1+\exp\left[\frac{(t- t_d)}{t_T}\right]},  
 \end{equation}
 where $n_c= \overline{ n}(t \rightarrow \infty)$ is the average number of Ar atoms in the cluster under condition of thermal equilibrium, and $\Delta(t,T)=$ $\frac{\epsilon(\overline{n}(t))  -\mu_{eff}(t)}{kT}$. The parameter $t_T$ represents the scaling time of cluster growth during the non-Maxwellian and quasi-equilibrium stages of nucleation. The scaling time-shift parameter $t_d$ takes into account a time-delay in formation of first tetrahedral shells and formally describes a motion of $ \mu_{eff}(t)$ towards the upper cluster shells with the atomic binding energy $\epsilon(n)$. The full derivation of the time-delay in the formation of ArH$^{+}$, Ar$_{2}$H$^{+}$, and Ar$_{3}$H$^{+}$ molecules needs to include an accurate analysis of few-body collisions, which are out of the scope of the simple two-state model. Thus, the simplified time-dependence of $\overline{n}(t)$ from Eq.2 cannot provide accurate initial conditions at t=0, but it describes well an evolution of $\overline{n}(t)$ for the entire time of simulations.
 
\begin{figure}[b]
\includegraphics[scale=0.325]{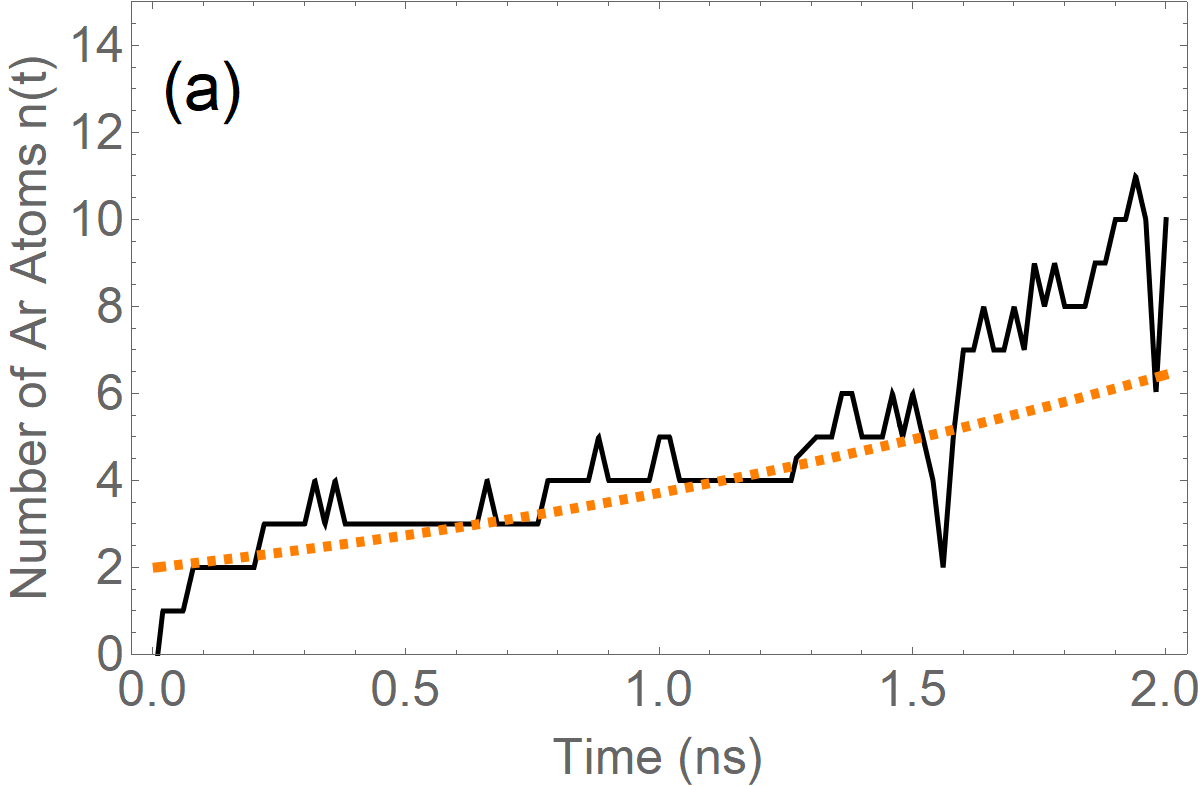} 
\includegraphics[scale=0.325]{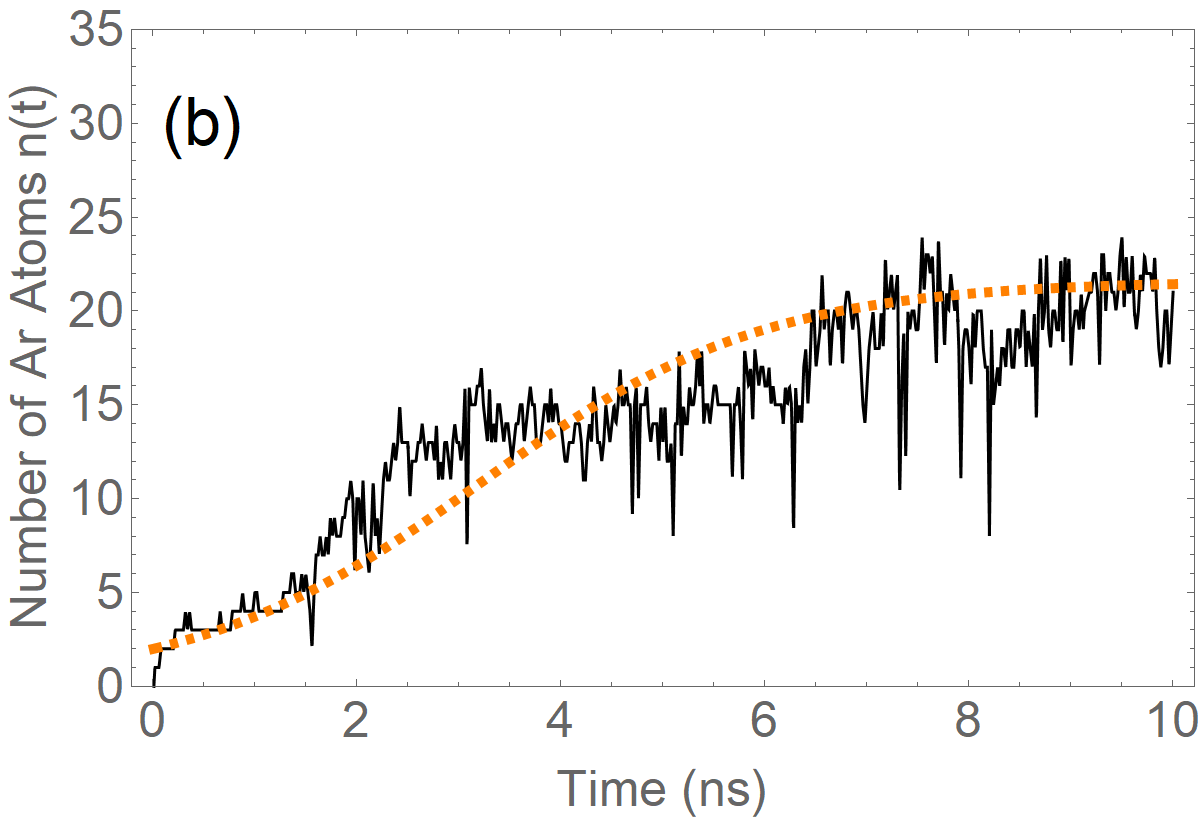}
\caption{\label{fig:Fluct_Onset_Single}The number of Ar atoms n(t) in the single $H^+$ ion cluster as a function of the time in the nonequilibrium stage of nucleation: (a) Formation of the deep shells in Ar$_n$H$^+$ clusters for the time interval t= 0 - 2 ns after creation of the H$^+$ ion in the argon bath gas; (b) The intermediate stage of Ar$_n$H$^+$ growth and development of the thermal fluctuations of $n(t)$ inside the time interval $0< t<$ 10 ns. The gas temperature and density are T= 90K and n$_g$= 10$^{20}$ cm$^{-3}$ respectively. Theoretical curve is computed using thermodynamic formula for the two state approximation given by Eq.2 with the kinetic parameters $t_d\simeq 3.2$ ns, $t_T= 1.4$ ns, and $n_c$=21.6. }
\end{figure}

 The reduction of the absolute value $|\mu_{eff}(t)|$ with time leads to the population of cluster shells with smaller binding energies $|\epsilon(n)|$ and thus stimulates an increase of the average cluster size $\overline{n}(t)$. At $t \rightarrow \infty$ the asymptotic value of $\mu_{eff}(t)$ matches to the equilibrium chemical potential $\mu(N,T)$: $\mu_{eff}(t) \rightarrow \mu(N,T) $, and $ \overline{n}(t\rightarrow \infty) \rightarrow n_c$. In Fig.\ref{fig:single100ns90k}, the value of $ \mu_{eff}(t)$ has been shifted up to energies $\epsilon(n)$ of Ar$_{n}$H$^{+}$ clusters with $n \sim 10-14$ for the time interval of few units of scaling time $t_d$, i.e approximately for 3$t_d \sim $10 ns. This reflect the process of the time-dependent population of the three deepest cluster shells of Ar atoms. From Eq.2 we can conclude, that at $t= t_d$ an average size $\overline{n}(t)$ has to be about 50\% of its thermal equilibrium value $n_c=21.6$ at T=90K. The values of kinetics parameters, $t_d\simeq 3.2$ ns and $t_T= 1.4$ ns, are inferred from the data depicted in Fig.\ref{fig:single100ns90k}. To illustrate the efficiency of the two state model, we show in Figs.\ref{fig:Fluct_Onset_Single}a and \ref{fig:Fluct_Onset_Single}b the initial stages of cluster growth for two time intervals $0 < t \lesssim 2$ ns and $0 < t \lesssim 10$ ns. 

 The nonequilibrium stage ($0 < t \lesssim 2$ ns) in Fig.\ref{fig:Fluct_Onset_Single}a does not indicate any detachments of Ar atoms from the first, deepest cluster shell. This nonequilibrium stage of growth cannot yield notable fluctuations of the cluster size $n(t)$ because the binding energies of Ar atoms in small clusters ($n \leq 8$) are significantly larger than the thermal energy at T=90K and thermal fluctuations cannot remove Ar atoms from these shells. Rare changes of $n(t)$ in Fig.\ref{fig:Fluct_Onset_Single}a are considered as attempts to increase the cluster size. Time of formation of the first two tetrahedral shells is estimated as 1 - 1.5 ns.
 The scaling value of thermalization time $t_T= $1.4 ns is in good agreement with our evaluation of the velocity relaxation time of Ar atoms. The $t_T$ value can be identified in Figs.\ref{fig:VasFig1}a, \ref{fig:VasFig1}b, \ref{fig:single100ns90k}, \ref{fig:Fluct_Onset_Single}a, and \ref{fig:Fluct_Onset_Single}b as the onset of the fast and strong up- and down- fluctuations of the cluster size n(t) with a simultaneous steady increase of the mean cluster size $\overline{n}(t)$. The quasi-equilibrium stage of cluster growth begins after 1.5 - 2 ns. The mean cluster size $\overline{n}(t)$ increases slowly until it shows a saturation at 10 - 15 ns since the beginning of the nucleation process, as is depicted in Figs.\ref{fig:single100ns90k} and \ref{fig:Fluct_Onset_Single}b. 
\section{\texorpdfstring{Equilibrium size distribution of Independent $Ar_nH^+$ clusters and cluster fluctuations}{Section: Equilibrium Size Distribution and Fluctuations}}
The size distribution of solid or liquid nano-particles is a fundamental characteristic required for analysis and modeling of many astrophysical and atmospheric phenomena. The size-distribution of small clusters depends on inter-atomic interactions and the nucleation kinetics. The attachment and detachment of atoms or molecules create time-dependent fluctuations of the cluster size even under the thermal equilibrium condition. Detailed analysis of these fluctuations can provide information on the cluster size distribution in an ensemble of independent particles. 
\subsection{Fluctuations of cluster size under the thermal equilibrium condition}
\begin{figure}
\includegraphics[scale=0.4]{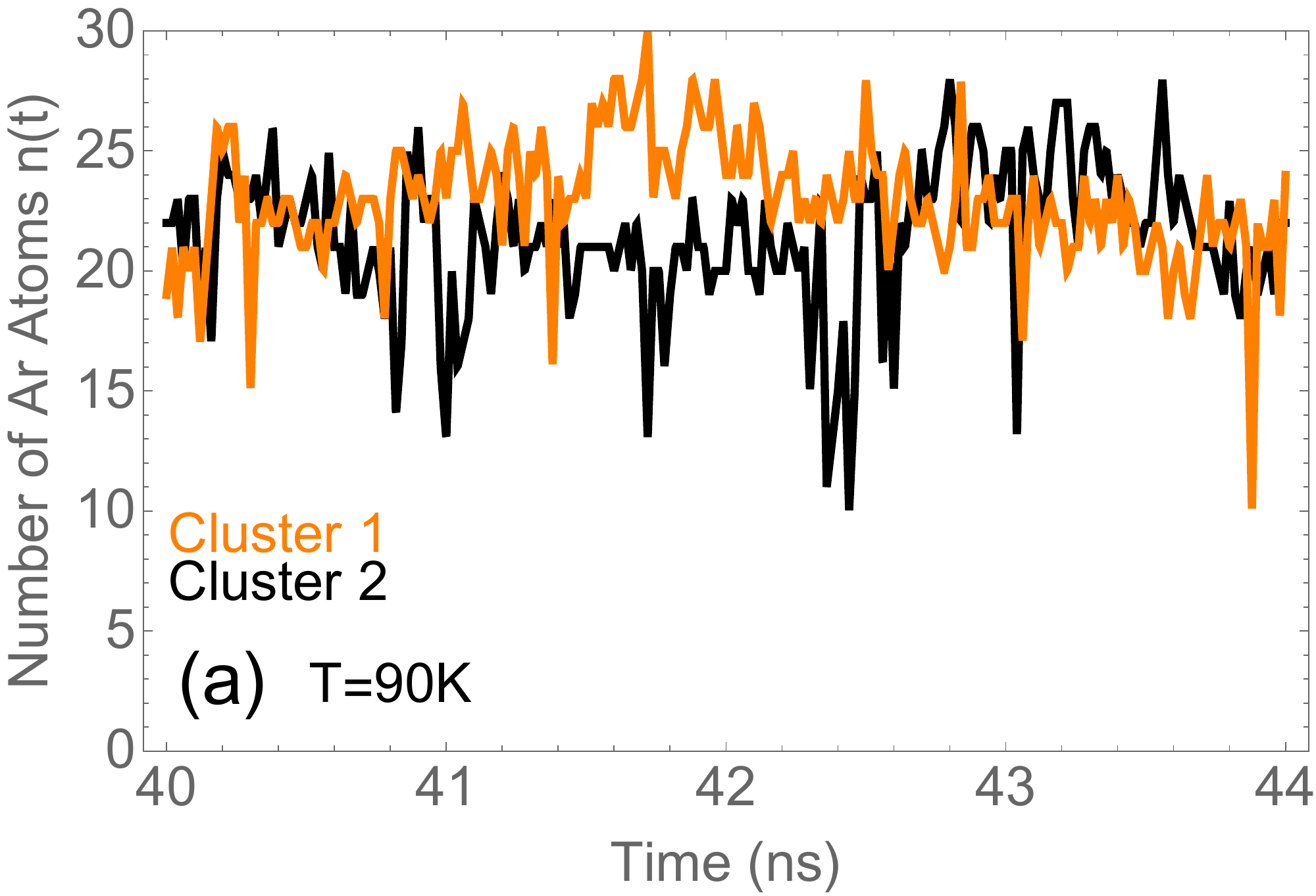} 
\includegraphics[scale=0.4]{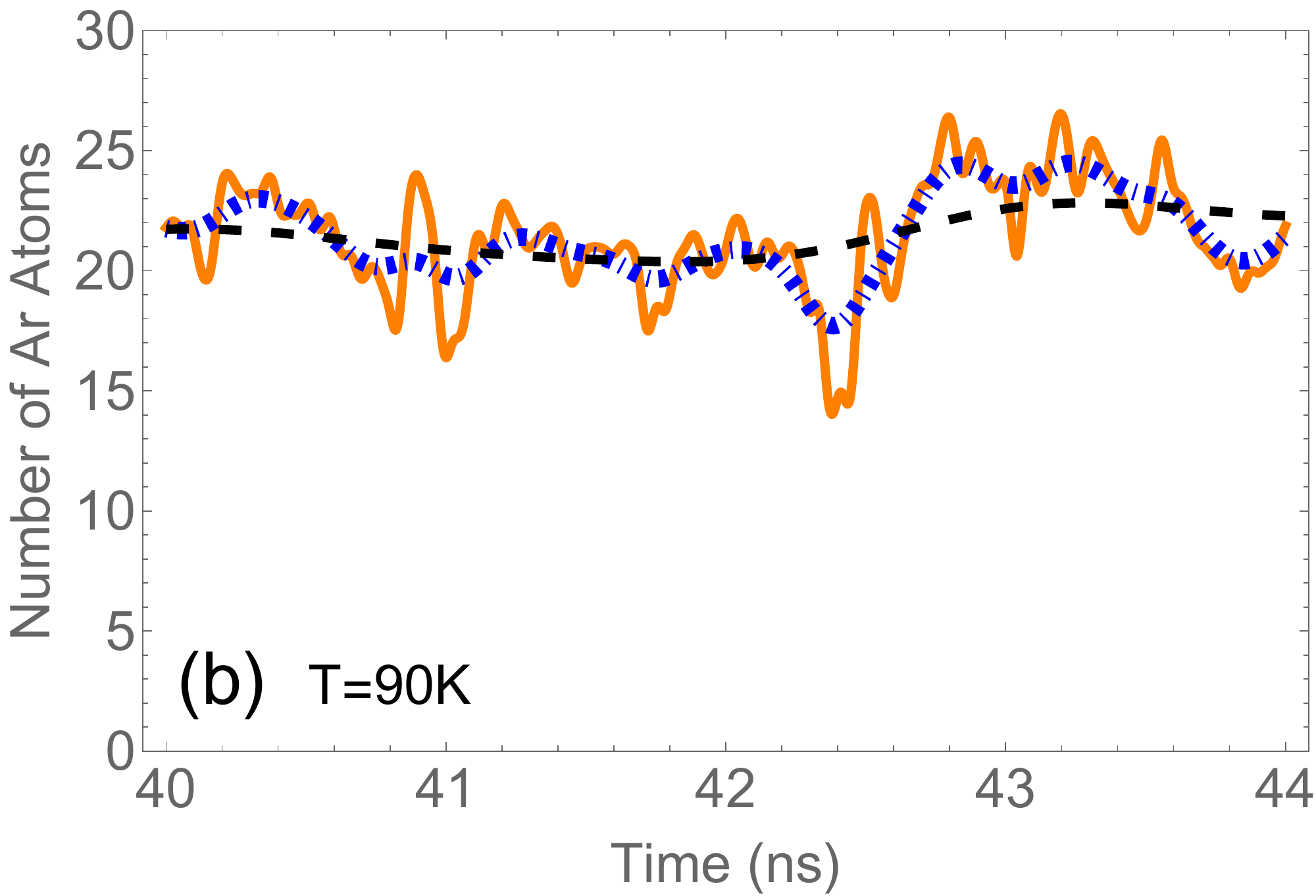}
\caption{\label{fig:Fluct_2Proton-TrajectoryN_ar_Equil.pdf} (a) The number of Ar atoms n$_1$(t) and n$_2$(t) in two independent clusters, 1 and 2, are shown as functions of the time $t$ under the thermal equilibrium conditions inside the time interval t= 40 ns - 44 ns. The gas temperature and density are T= 90K and n$_g$= 10$^{20}$cm$^{-3}$ respectively. (b) The time-scale filtering of fluctuations are shown in Fig.7b; the smoothed curves in Fig.7b are shown for the cluster 2 from Fig.7a. The averaging time for the Gaussian filters are $2\sigma=$ 40 ps, 0.2 ns, and 1 ns for solid, dot-dashed, dashed curves respectively.}
\end{figure}

The "flat" behavior of $\overline{n}(t)$ at large time $t$ and intensive fluctuation regime are indicators of the thermal equilibrium between "cluster-bound" and free Ar atoms. The number of Ar atoms $n(t)$ in a Ar$_n$H$^+$ cluster are shown in Fig.\ref{fig:Fluct_2Proton-TrajectoryN_ar_Equil.pdf}a for two clusters formed by two independent protons under the thermal equilibrium conditions. Two "trajectories" of MD simulations n$_1$(t) and n$_2$(t) for this independent clusters are shown as functions of time inside the time interval between 40 ns and 44 ns. Cluster size fluctuations are characterized by different time scales and amplitudes. Averaging of $n(t)$-function over different time-intervals shows separate typical frequency and amplitudes of cluster size-fluctuations. For example, the results of averaging (filtering) of the cluster size fluctuations are shown in Fig.\ref{fig:Fluct_2Proton-TrajectoryN_ar_Equil.pdf}b for different Gaussian filtering intervals $2\sigma=$ 40 ps, 0.2 ns, and 1 ns. Fast fluctuations correspond to the thermal attachment/detachment of Ar atoms from cluster shells. Slow fluctuations can be attributed to a long-term relaxation of thermo-dynamical parameters of the gas environment around clusters. The long-term fluctuations exist for all Ar$_n$H$^+$ clusters and they have similar scale of typical fluctuations, but happen at different times and cluster locations.

The fluctuation pattern of the cluster size-distribution $N_c(n,t)$ has also been found in simulations of cluster growth for an ensemble of H$^+$ ions embedded into Ar gas (Figs.\ref{fig:VasFig1}a and \ref{fig:VasFig1}b).
 Different stages of nucleation processes shown in Figs.\ref{fig:VasFig1}a and \ref{fig:VasFig1}b for the ensemble of 200 H$^+$ ions and thermal fluctuations of the cluster population can be investigated in detail using data of the cluster formation by individual H$^+$ ions. 
Strong time-fluctuations of n(t) near the average equilibrium value illustrate the diffusive nature of cluster growth under the thermal equilibrium conditions. The scale of cluster fluctuations shown in Figs.\ref{fig:Fluct_2Proton-TrajectoryN_ar_Equil.pdf}a and \ref{fig:Fluct_2Proton-TrajectoryN_ar_Equil.pdf}b provides accurate information on the stationary equilibrium distribution of Ar$_n$H$^+$ cluster sizes at different temperatures.

 \subsection{\texorpdfstring{Equilibrium size-distribution of Ar$_n$H$^+$ clusters}{Subsection: Equilibrium Size distribution}}
 Parameters of the time evolution of the average cluster size $\overline{n}(t)$ and time-dependent fluctuations, n(t)- $\overline{n}(t)$, around the average value yield unique information on the cluster size distribution during the quasi-equilibrium and equilibrium stages of the cluster growth. Fluctuations can be considered as the bases for the diffusion process in the parametric $\{n\}$-space of the Ar$_{n}$H$^{+}$ clusters \cite{Zeldovich,LLkinetics, ReviewCNT1,ReviewCNT2}. Exchange of Ar atoms between cluster shells and free Ar gas during the quasi-equilibrium or equilibrium stages of nucleation is an example of a random walk in the space of cluster size $n$. 
 
 The diffusion characteristics of the Ar$_n$H$^+$ growth and the stationary cluster size distributions $P(n,T)$ have been inferred from the results of MD simulations at different temperatures and ion concentrations. The empirical probabilities to detect Ar$_{n}$H$^{+}$ cluster with $n$ Ar atoms during specified time interval $\tau $ is given by the normalized Probability Distribution Function (PDF) $P(n,t,\tau,T)$. The empirical probability $P(n,t,\tau,T)= N(n,t,\tau,T)/N(t,\tau,T)$ is defined as a ratio of the number of $N(n,t,\tau,T)$ realization of Ar$_{n}$H$^+$ clusters with $n$ atoms for the time interval $\tau$ to the total number of realization $N(t,\tau,T) =\sum_{n}N(n,t,\tau,T)$. The time-interval $\tau$ has been selected to be high enough for an accurate statistical evaluation of $n$-distributions. The defined PDF does not depend on the time, under condition of the thermal equilibrium and describes stationary thermal probabilities $P(n,T)$.
 
\begin{figure}
\includegraphics[scale=0.35]{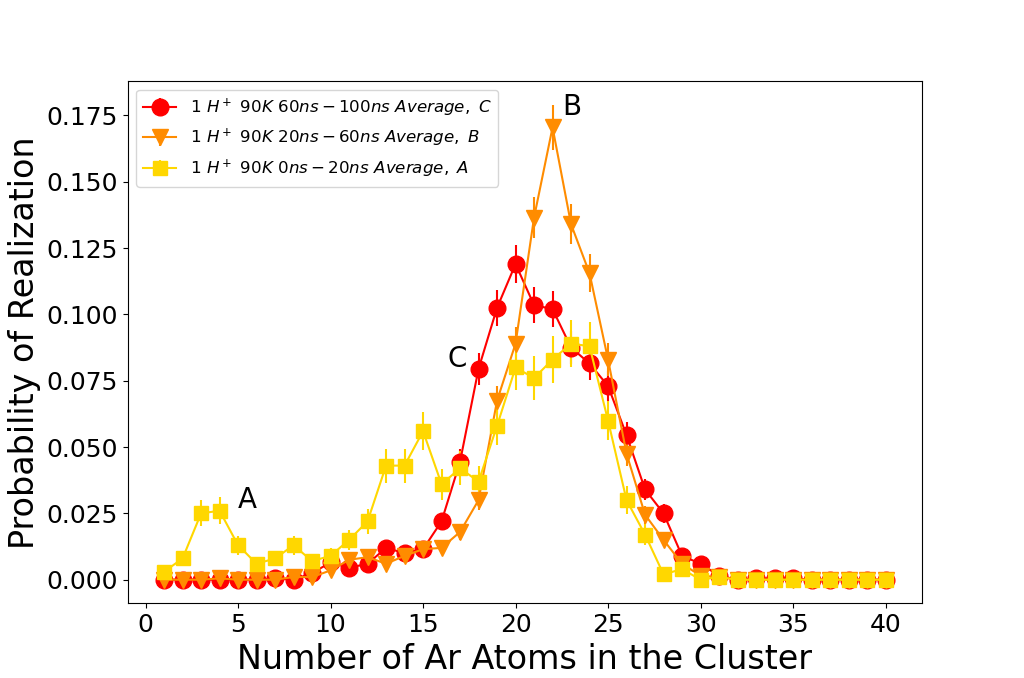}
\caption{\label{fig:vasFig2}The empirical probability $P(n,t,\tau,T) $ to find the Ar$_{n}$H$^{+}$ clusters with $n$ Ar atoms in the cluster shells. The total simulation time t= 100 ns for a single H$^+$ ion at T=90K. The curves A, B, and C are shown at different times: curve A, the earlier stage of cluster formation 0$\leq t \leq $20 ns, curve B just after thermalization 20$\leq t \leq $60 ns, and curve C 60$\leq t \leq $100 ns at the end of the MD simulation process. The averaging interval for the later distributions is $\tau $ =40 ns, and for the early stage A is 20 ns. The 0-20 ns time interval includes essentially nonequilibrium stage of the cluster nucleation and this causes a visible fraction of small clusters in the empirical probability given by the curve A.}
\end{figure}

 We have computed $P(n,t,\tau,T)$ at different time $t$ to investigate the kinetics of the independent cluster growth under quasi-equilibrium and equilibrium conditions. Results are shown by curves A, B, and C in Fig.\ref{fig:vasFig2}. The error bars in Fig.\ref{fig:vasFig2} indicate a standard deviation of the MD simulation results.
Cluster size distributions at different times $t$ reflect different phases of the cluster shell formation. A long tail of small clusters with $n \leq 16$ as is shown in Fig.\ref{fig:vasFig2} (curve A), is formed during earlier stages of cluster growth under nonequilibrium conditions and this is clearly reflected in $P(n,t,\tau=20 ns,T)$ for $t\leq 20$. More detailed information about the cluster $n$-distributions at short time $t\leq 20$ ns requires significantly shorter intervals $\tau$ due to the fast non-thermal growth of clusters (Fig.\ref{fig:single100ns90k}) and so, an increasing number of independent simulating trajectories $\{n(t)\}$.

The quasi-equilibrium and equilibrium stages of the cluster formation arise after $t\geq 20$ ns (B and C curves in Fig.\ref{fig:vasFig2}). They show broad distributions of the cluster sizes $n$ between 15 and 30 Ar atoms. These probabilities are controlled by the distribution of the cluster Ar-detachment energies $u(n)$, which have deeper local minima near $n \sim$ 19-23 as it is shown in Fig.\ref{fig:ChemPotentials_Misha}. The low temperature T=90K allows Ar atoms to occupy outer cluster shells. These shells intensively exchange by Ar atoms with the thermal Ar gas creating large $n$ fluctuations (Fig.\ref{fig:single100ns90k}) with the standard deviation $\sigma(n,T)\simeq$ 4 Argon atoms. The value of $\sigma(n,t)$, inferred from the analysis of time-dependent fluctuations at the thermal equilibrium stage, allows to compute the theoretical value of the full width at half maximum (FWHM) of the thermal distribution: FWHM$ \simeq$ 2.34 $\sigma(n,t) \simeq 9.3$. This matches well to the simple estimation of the FWHM$_C$ value of the thermal distribution P(n,T) (curve C in Fig.\ref{fig:vasFig2}): FWHM$_{C}\simeq$10, if the n-distribution of the C-curve is approximated by a Gaussian. The peak of $P(n,t,\tau,T)= P(n,T)$ around the minimal values of $u(n)$ at the region $n\sim 19 - 23$ and strong n-fluctuations are mostly formed at $t > 20$ ns.\\

\section{\texorpdfstring{Kinetics of the cluster formation for ensemble of H$^+$ ions}{Section: H Ion ensemble}}
The kinetics of cluster nucleation and their thermal equilibrium parameters can be modified by the mutual influence of Ar$_{n}$H$^{+}$ in the cluster formation process. At high concentration of H$^+$ ions, the capture of free Ar atoms by different clusters becomes a competitive process. Additional complexity of nucleation arises at low gas temperature, when strong correlations between Ar$_{n}$H$^{+}$ clusters lead to their aggregation to a new phase, the large scale Ar$_n$H$^+_m$ droplet or nano-crystal \cite{Oliver}. This transition may occur via several paths, such as consolidation of strongly-bound Ar$_{4}$H$^{+}$ clusters or as a coalescence process of absorption of small and medium clusters by larger nano-particles \cite{LLkinetics}. We have simulated the cluster nucleation using different initial ensembles of H$^+$ ions, N$_{H^+}$ from 1 to 200 H$^+$ ions at the temperatures T=90K and 200K.

\begin{figure}
\includegraphics[scale=0.35]{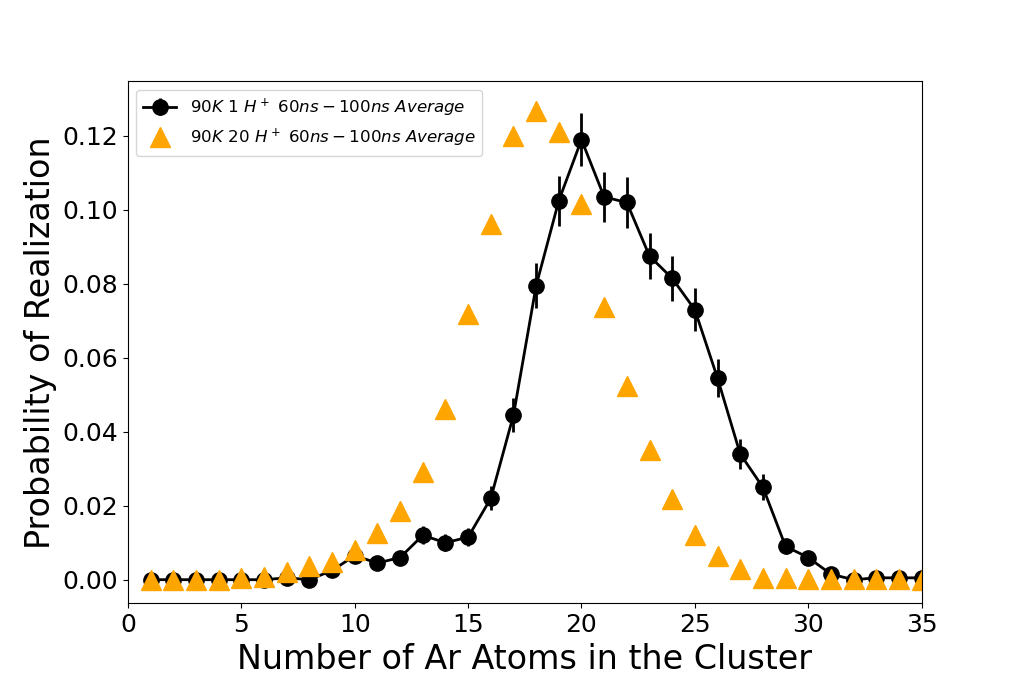}
\caption{\label{fig:vasFig3}The probability distribution function $P(n,N_{H^+})$ to find Ar$_{n}$H$^{+}$ clusters with $n$ Ar atoms in the cluster shells as a function the number of argon atoms $n$. Ar$_{n}$H$^{+}$ clusters are formed by the initial ensemble of N$_{H^+}$ ions. The total simulation time is t= 100 ns. The gas temperature T= 90 K, with either N$_{H^+}$=1 (circles), or N$_{H^+}$=20 (triangles) H$^+$ ions, all at Ar density $10^{20}$ cm$^{-3}$. Data is averaged over the time interval 60-100 ns, error bars for the 20 H$^+$ ion curve are on the order of the marker size. The introduction of more ions leads to a reduction of the mean cluster size compare with the independent proton growth.}
\end{figure}

To clarify the influence of an increasing H$^+$ density on the nucleation process, we have performed simulations of the Ar$_n$H$^+$ nucleation at N$_{H^+}$=1 and 20 H$^+$ ions at the temperature T=90K and the constant number of Ar atoms N$_{Ar}$=10$^3$ in the simulation box. Results are shown in Fig.\ref{fig:vasFig3}. The significant increase of the number of H$^+$ ions (triangles in Fig.\ref{fig:vasFig3} for N$_{H^+}$= 20) leads to a reduction of the mean cluster size compared with the independent proton growth (circles).
At T=90K, the 20 H$^+$ ions have captured about 36\% of free argon atoms. The density of the free Ar gas became around 64\% of the initial gas density and that shifts the negative chemical potential of free Ar atoms down towards Ar cluster binding energies on the value $\Delta \mu_{gas}(T)\simeq \ln[0.64]$ kT $\simeq$ - 0.45 kT. Argon atoms from the largest size clusters are easily evaporated due to such shifts. In this simplified estimations we have neglected changes of the mean field potential. 

\begin{figure}
\includegraphics[scale=0.35]{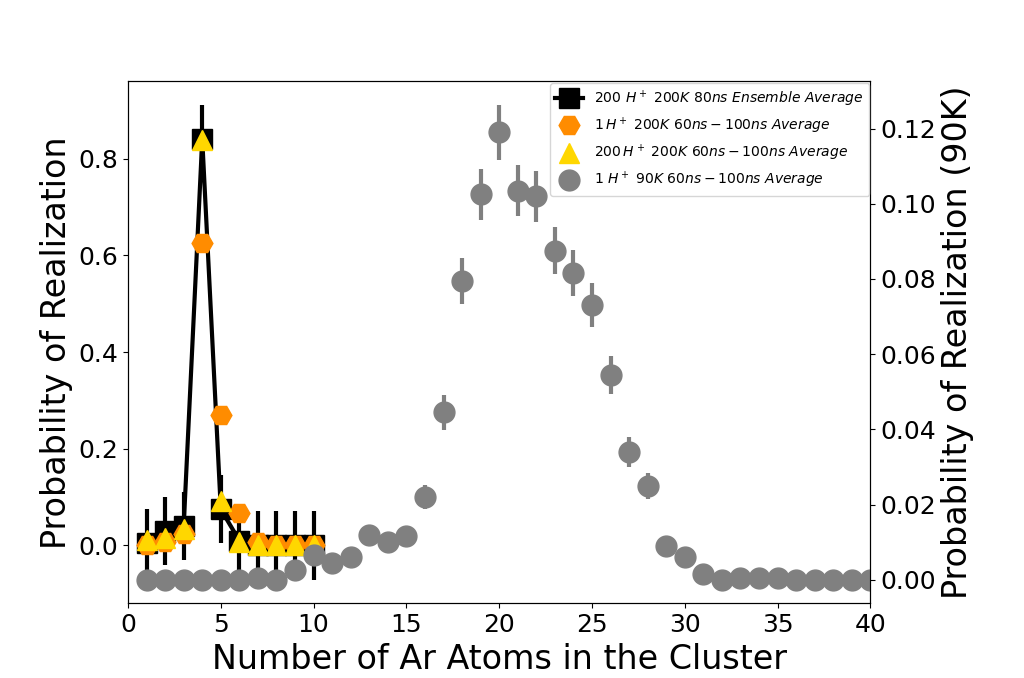}
\caption{\label{fig:VasFig4}The thermal equilibrium probability $P(n,N_{H^+},T)$ to detect the Ar$_{n}$H$^{+}$ clusters with the $n$ Ar atoms for the ensembles of N$_{H^+}$= 1 and 200 H$^+$ ions. The results of the independent H$^+$ ion growth are show for T=200K (hexagons, left vertical axis) and T= 90K (circles, right vertical axis). The cluster size distribution $P(n,N_{H^+}=200,T=200K)$ for the averaged value over the ensemble of 200 H$^+$ ions at the single snapshot at $t$=80 ns are shown (squares, left vertical axis) with corresponding error bars. The triangles (left vertical axis) show results of $P(n,N_{H^+}=200,T=200K)$ obtained with time and ensemble averaging. The error bars for this case is smaller than the marker size. }
\end{figure}

The cluster size distribution for the ensemble of 200 H$^+$ ions has been inferred from our MD simulations at the temperature T=200K and results are shown in Fig.\ref{fig:VasFig4} by the triangles obtained with time and ensemble averaging. 
The squares with corresponding error bars shows the same data for T=200K averaged only over the ensemble of 200 H$^+$ at the single snapshot of $t$=80 ns. This narrow size-distribution is peaked around Ar$_4$H$^+$ clusters. Clusters with large atomic shells cannot be observed at high temperatures, because thermal fluctuations destroy these shells. For example, at the temperature T=200K only small cluster with the typical size between 2-7 Ar atoms in the shells can be stable. For comparisons between multi-proton and single proton nucleation, the equilibrium distribution function of the cluster size $P(n,T=200K)$ has been inferred from the single H$^+$ ion simulations of the independent cluster growth at T= 200K and the results are shown in Fig.\ref{fig:VasFig4} with red down-triangles. The narrow $P(n,T=200K)$ distribution is sharply peaked at $n$=4 (the tetrahedral clusters) in contrast with the low temperature $P(n,T=90K)$, which is favorable to a broad range of larger cluster size with $n\sim$ 15-30. For comparison, the low temperature $n$-distribution of independent clusters at T=90K is also shown in Fig.\ref{fig:VasFig4} by circles.

The similarity of the high-temperature distributions for the single H$^+$ and 200 H$^+$ ions at T= 200K shows that although the cluster chemical potentials are sensitive to both the density of the H+ ions and Ar atoms; the large value of the Ar binding energy of the Ar$_4$H$^+$ cluster causes the cluster size distribution to be weakly sensitive to the proton density in the specific interval of temperature around 200K considered in the article. Therefore, cluster growth at chosen parameters can be seen mostly as a nucleation kinetics of independent small size clusters. This statement is strongly supported by results of time dependent kinetics of cluster nucleation in the ensemble of 200 seed $H^+$ ions previously shown in Fig.\ref{fig:VasFig1}a and \ref{fig:VasFig1}b. The abundances (n-distribution) of all small clusters with sizes $n\leq$10 have reached their equilibrium values around 2 ns. However, the quasi-equilibrium stage of nucleation (the orange circles and gold upside-down triangles (curve  in Fig.\ref{fig:VasFig1}a for the clusters Ar$_4$H$^+$ and Ar$_5$H$^+$respectively) shows the steady growth of stable tetrahedral clusters (the circles in Fig.\ref{fig:VasFig1}a). The development of thermal fluctuations is clearly seen in Fig.\ref{fig:VasFig1}a for Ar$_4$H$^+$ and Ar$_5$H$^+$ clusters. The anti-phase of Ar$_4$H$^+$ and Ar$_5$H$^+$ fluctuations results in an effective single Ar atom exchange between the tetrahedral clusters and free Ar gas. Any capture of free Ar atoms by Ar$_4$H$^+$ leads to formation of new Ar$_5$H$^+$ clusters and to reduction of Ar$_4$H$^+$. The thermal equilibrium detachment of a single Ar atom from Ar$_5$H$^+$ produces a new stable tetrahedral cluster.
The thermal energy fluctuation cannot destroy the deepest shell even at T=200 K. The Ar$_4$H$^+$ are most stable and abundant clusters for both the single- and multi-proton nucleation process as it is shown in Fig.\ref{fig:VasFig4}.
\section{Conclusions}
 Our analysis of MD simulations provide a consistent scenario for a new nucleation phase initiated by ions in neutral gases. The growth of ion clusters occurs via the formation of atomic shells around the ion seed particle. MD simulations show evidence for three distinct stages of cluster nucleation: nonequilibrium, quasi-equilibrium, and equilibrium stages. In the first nonequilibrium stage, the strong ion field removes the barriers to nucleation; thus, the probability to capture a gas atom into the deeply cluster-bound states (1T and 2T) is significantly higher than the probability for all detachment processes. This leads to the build up of inner shells during this stage. When a gas atom is captured into inner shells, a local release of high kinetic energies occurs. This causes the energy distributions of the atomic particles to become non-Maxwelllian. Once these energy distributions relax to a Maxwellian like distribution; the system moves into the second quasi-equilibrium stage. During this stage the equilibrium between the new phase (clusters) and the free gas has yet to be established. Thus, we see a steady growth of the cluster size and number of clusters in multi-ion systems, up to the final thermal equilibrium stage. The notable feature of the quasi-equilibrium stage is the onset and evolution of fast and strong fluctuations of the atom number. The size of the cluster increases with simultaneous decrease of binding energies in the outer cluster shells. This stimulates detachment processes and the exchange of atoms between the clusters and the free gas. Then system moves into the final thermal-equilibrium, stage, where the cluster size distribution reaches a steady state value. The density of gas and ions, temperature, and parameters of cluster shell structures regulate the cluster size-distribution. The time-dependent fluctuations of the cluster size can be used to predict parameters of the steady state distribution, specifically, the cluster size dispersion in an ensemble of growing clusters.
 
 Nucleation of nano-particles in astrophysical and planetary environments include more complicated physics of the cluster formation in molecular gases, such as  N$_2$, CO$_2$, H$_2$O, CH$_4$  and others; due to the internal molecular structure and more complicated intermolecular and intramolecular interactions. We intend to investigate kinetics of nano-cluster formation in important astrophysical gases and apply new theoretical parameters of nucleation for modeling observational data.  Our results, reported in this article, will be useful in analysis of the cluster growth rates and size distributions of charge-seeded clusters as function of the gas parameters and molecular structures.

\section*{Acknowledgments}
 The work of MB, JS, and RC on this project was supported via the UCLA grant ONRBAA13-022, HRS and VK acknowledge support from the NSF through a grant for ITAMP at Center for Astrophysics $|$ Harvard \& Smithsonian. One of the authors (DV) is also grateful for the support received from the National Science Foundation through grants PHY-1831977 and HRD-1829184.

\bibliography{references}
\end{document}